\documentclass[12pt]{article} 

\begin{document}

\baselineskip=12pt plus 1pt minus 1pt 

\begin{center}

{\large \bf Molecular Spectra from Rotationally Invariant Hamiltonians \\
Based on the Quantum Algebra su$_q$(2) \\ and Irreducible Tensor Operators 
under su$_q$(2)}  

\bigskip\bigskip\bigskip 

{DENNIS BONATSOS$^{\#}$\footnote{e-mail: bonat@inp.demokritos.gr}, 
B. A. KOTSOS$^*$\footnote{e-mail: bkotsos@teilam.gr}, 
P. P. RAYCHEV$^\dagger$\footnote{e-mail: raychev@phys.uni-sofia.bg, 
raychev@inrne.bas.bg}, 
P. A. TERZIEV$^\dagger$\footnote{e-mail: terziev@inrne.bas.bg} }

\bigskip 
{$^{\#}$ Institute of Nuclear Physics, N.C.S.R. ``Demokritos'', \\ GR-15310 
Aghia Paraskevi, Attiki, Greece}

\medskip
{$^*$ Department of Electronics, Technological Education Institute, 
\\ GR-35100 Lamia, Greece} 

\medskip
{$^\dagger$ Institute for Nuclear Research and Nuclear Energy, Bulgarian
Academy of Sciences, \\ 72 Tzarigrad Road, BG-1784 Sofia, Bulgaria}

\bigskip\bigskip 

\centerline{\bf ABSTRACT} 

\end{center} 

The rotational invariance under the usual physical angular momentum of the 
su$_q$(2) Hamiltonian for the description of rotational molecular spectra 
is explicitly proved and a connection of this Hamiltonian to the formalism 
of Amal'sky is provided. In addition, a new Hamiltonian for rotational spectra
is introduced, based on the construction of irreducible tensor operators 
(ITOs) under su$_q$(2) and use of $q$-deformed tensor products 
and $q$-deformed Clebsch--Gordan coefficients. The rotational invariance 
of this su$_q$(2) ITO Hamiltonian under the usual physical angular momentum 
is explicitly proved and a simple closed expression for its energy spectrum 
(the ``hyperbolic tangent formula'') is introduced. Numerical tests against 
an experimental rotational band of HF are provided.  

\bigskip\bigskip

\section{Introduction}

Quantum algebras \cite{Chari,Bieden,Klimyk} 
have started finding applications in the description 
of symmetries of physical systems over the last years \cite{PPNP},
triggered by the introduction of the $q$-deformed harmonic oscillator 
\cite{Bie,Mac,SunFu}.  
In one of the earliest 
attempts, a Hamiltonian proportional to the second order Casimir operator 
of su$_q$(2) has been used for the description of rotational molecular
\cite{CPL175} and nuclear 
spectra \cite{RRS} and its relation to the Variable Moment of Inertia Model 
\cite{PLB251} has been clarified. 

However, several open problems remained:

a) Is the su$_q$(2) Hamiltonian invariant under the usual su(2) Lie 
algebra, i.e.  under usual angular momentum, or it breaks spherical 
symmetry and/or the isotropy of space? 

b) How does the physical angular momentum appear in the framework 
of su$_q$(2)? Is there any relation between the generators of su$_q$(2) 
and the usual physical angular momentum operators? 

c) How can one add angular momenta in the su$_q$(2) framework? In other words,
how does angular momentum conservation work in the su$_q$(2) framework? 

Answers to these questions are provided in the present paper, along with 
connections of the su$_q$(2) model to other formalisms. 

In Section 2 a brief description of the $q$-deformed harmonic oscillator is 
given, while in Section 3 the connection between the $q$-deformed 
harmonic oscillator and the usual harmonic oscillator is established 
through the use of $q$-deforming boson functionals, a tool which will
be needed in the framework of su$_q$(2) as well. The su$_q$(2) 
formalism is briefly described in Section 4, while in Section 5 a
representation of su$_q$(2) in terms of $q$-deformed boson operators 
is given. Based on this boson representation the connection between the 
su$_q$(2) quantum algebra and the usual su(2) Lie algebra is established 
in Section 6 and used in Section 7 for proving  
explicitly that the su$_q$(2) Hamiltonian does 
commute with the generators of su(2), i.e. with the generators of usual 
physical angular momentum. Therefore the su$_q$(2) Hamiltonian does not 
violate the isotropy of space and does not destroy spherical symmetry. 
The su$_q$(2) basis turns out to be identical to the usual su(2) basis 
in the boson representation under discussion. 
In addition, it turns out that the angular momentum quantum numbers 
appearing in the description of the su$_q$(2) states are exactly the same 
as the ones appearing in the states
of su(2), establishing an one-to-one correspondence between the two sets of
states (in the generic case in which the deformation parameter $q$ is not 
a root of unity). 

Taking advantage of the results of Section 7, we write in Section 8 the 
eigenvalues of the su$_q$(2) Hamiltonian as an exact power series in 
$l(l+1)$ (where $l$ is the usual physical angular momentum). An approximation 
to this expansion, studied in Section 9, leads to a closed energy formula for 
rotational spectra introduced by Amal'sky \cite{Amal}. 

We then turn in Section 10 into the study of irreducible tensor operators 
under su$_q$(2) \cite{STK593,STK690}, 
constructing the irreducible tensor operator of rank one 
corresponding to the su$_q$(2) generators. We also define tensor products 
in the su$_q$(2) framework and construct the scalar square of the angular 
momentum operator, a task requiring the use of $q$-deformed Clebsch--Gordan 
coefficients \cite{STK593}. In addition to exhibiting explicitly how addition 
of angular momenta works in the su$_q$(2) framework, this exercise leads 
to a Hamiltonian built out of the components of the above mentioned 
irreducible tensor operator (ITO), which can also be applied to the 
description of rotational spectra. We are going to refer to this Hamiltonian 
as the {\sl  su$_q$(2) ITO Hamiltonian}. 

The fact that the su$_q$(2) ITO Hamiltonian does commute with the generators 
of the usual su(2) algebra is shown explicitly in Section 11. Based on the 
results of Section 11, we express in Section 12 the eigenvalues 
of the su$_q$(2) ITO Hamiltonian 
as an exact power series in $l(l+1)$, where $l$ is the 
usual physical angular momentum. An approximation to this series, 
studied in Section 13, leads to a simple closed formula for the spectrum
(the ``hyperbolic tangent formula''). 

Finally in Section 14 all the exact and closed approximate energy formulae 
obtained above are compared to the experimental spectrum of a rotational 
band of HF, which exhibits sizeable deviations from pure rotational behavior,
as well as to the results provided by the usual 
rotational expansion and by the Holmberg--Lipas formula \cite{Lipas}, which 
is probably the best two-parameter formula for the description 
of rotational nuclear spectra \cite{Casten}. A discussion of the present 
results and plans for future work are given in Section 15. 

\section{The $q$-Deformed Harmonic Oscillator}

The $q$-deformed harmonic oscillator \cite{Bie,Mac,SunFu} is defined 
by means of the operators  $a^\dagger$ and $ a$, which 
are referred to as $q$-deformed boson creation and annihilation operators, 
together with the $q$-deformed number operator $N$. 
These operators close the $q$-deformed Heisenberg--Weyl algebra
\begin{equation}\label{eq:q1} 
[a, a^\dagger ]_q \equiv a a^\dagger-
q a^\dagger a = q^{-N}, \qquad 
[a, a^\dagger ]_{q^{-1}} \equiv a a^\dagger -
q^{-1} a^\dagger  a = q^N, 
\end{equation}
\begin{equation}\label{eq:q2} 
[ N, a^\dagger ] = a^\dagger,  \qquad
[N, a]=-a. 
\end{equation}
From Eq. (\ref{eq:q1}) one can easily see that the 
products of operators $a$ and $a^\dagger$ are equal to
\begin{equation}\label{eq:q3}
a^\dagger a =[N], \qquad a a^\dagger = [N+1], 
\end{equation}
where the square brackets denote $q$-operators, defined as 
\begin{equation}\label{eq:q4}
\left[X \right] = \frac{q^X-q^{-X}}{q-q^{-1}}. 
\end{equation}
$q$-numbers are also defined in the same way. 
For the deformation parameter $q$ two different cases occur:

a) If $q$ is real, one can write $q=e^\tau$, where $\tau$ is real. 
Then one immediately sees that
\begin{equation}\label{eq:q5}
[X] = {\sinh \tau X \over \sinh \tau}.
\end{equation}

b) If $q$ is a phase factor (but not a root of unity, in which case 
one has $q^n=1$, with $n\in {\bf N}$), one can write 
$q=e^{i\tau}$, where $\tau$ is real. Then one has 
\begin{equation}\label{eq:q6}
[X]= {\sin \tau X \over \sin \tau}. 
\end{equation}

In both cases one has 
\begin{equation}\label{eq:q7} 
[X] \to X  \quad {\rm as} \quad q\to 1. 
\end{equation}

There exists a state $\vert 0 \rangle_q$ (called the $q$-vacuum) with the
properties
\begin{equation}\label{eq:q8}
a \vert 0 \rangle_q=0 , \qquad N \vert  0 \rangle_q=0 .
\end{equation}
Acting on this state with the operator $ a^\dagger$ repeatedly, one can 
build up the states
\begin{equation}\label{eq:q9}
\vert n \rangle_q = \frac{(a^\dagger)^n}{\sqrt{[n]!}}  \vert 0
  \rangle_q ,
\end{equation}
where the quantity 
\begin{equation}\label{eq:q10} 
[n]! =[n]\,[n-1] \ldots [1] 
\end{equation} 
is the $q$-factorial.

The states (\ref{eq:q9}) form an orthonormal basis 
\begin{equation} \label{eq:q11}
 _q\langle n \vert m \rangle_q = \delta_{nm} , 
\end{equation} 
and are eigenvectors of the $q$-deformed number operator
\begin{equation}\label{eq:q12}
N \,\vert \, n \rangle_q = n\, \vert \, n \rangle_q. 
\end{equation}
From the last equation it is clear why the states of Eq. (\ref{eq:q9}) 
are interpreted as states containing $n$ $q$-deformed bosons. 

\section{Connection of the $q$-Deformed Oscillator to the \hfill\break 
Usual Harmonic Oscillator} 

In the limit $q \to 1$ (or $\tau \to 0$) Eq. (\ref{eq:q1}) 
is reduced into the commutation relation for
usual  boson creation and annihilation operators, for which 
the symbols  $b^\dagger$ and $b$ will be used 
\begin{equation}\label{eq:q13}
\left[ b, b^\dagger \right] = 1.
\end{equation}
Furthermore, in the limit $q\to 1$ Eq. (\ref{eq:q2}) is reduced into the form
\begin{equation}\label{eq:q14}
 [B,b^\dagger] = b^\dagger, \qquad [B, b]= -b ,
\end{equation}
where by $B$ we denote the number operator, for which the following analogues 
of Eq. (\ref{eq:q3}) are valid
\begin{equation}\label{eq:q15}
b^\dagger b=B, \qquad b b^\dagger =B+1.  
\end{equation}

In analogy with Eq. (\ref{eq:q8}) one can introduce the vacuum state $\vert
 0 \rangle_c$ (the ``classical'' vacuum) with the properties  
\begin{equation}\label{eq:q16}
b \vert  0 \rangle_c =0, \qquad B\vert 0\rangle_c =0.  
\end{equation}
Then one can build up the states
\begin{equation}\label{eq:q17}
  \vert n \rangle_c = \frac{(b^\dagger)^n}{{\sqrt n!}} \vert 0  \rangle_c,
\end{equation}
which form an orthonormal basis 
\begin{equation}\label{eq:q18} 
_c\langle n \vert m \rangle_c = \delta_{nm},
\end{equation}
and are eigenstates of number operator $B$, i.e.  
\begin{equation}\label{eq:q19}
B \vert n \rangle_c = n \vert n \rangle_c. 
\end{equation} 
From the last equation it is clear that the states of Eq. (\ref{eq:q17}) 
are states which contain $n$ bosons. 
The analogy between the ``standard'' and the $q$-deformed case is clear.

There is however a closer relation between the operators
$a^\dagger$ ($a$) and $b^{\dagger}$ ($b$). In the space
determined by the vectors of Eq. (\ref{eq:q17}) one can define the operators
\cite{Song} 
\begin{equation}\label{eq:q20}
a = b \sqrt{\frac{[B]}{B}} \qquad \mbox{and} \qquad a^\dagger =
\sqrt{\frac{[B]}{B}}\,b^\dagger. 
\end{equation}
It should be noticed that the definitions of Eq.  (\ref{eq:q20}) are
meaningful only in the space of vectors $\vert n \rangle_c$, i.e. 
when the operators $a^\dagger$ and $a$ act on the vectors of 
Eq. (\ref{eq:q17}). 
As a consequence the operator $B$, which appears in the denominator of 
Eq. (\ref{eq:q20}), can be replaced by the  
number  $n$ only when $a^\dagger$ or $a$ act on $\vert n \rangle_c$.

One can now 
calculate the bilinear products of $a^\dagger$ and $a$
\begin{equation}\label{eq:q21} 
a^\dagger a = \sqrt{\frac{[B]}{B}} \underbrace{ b^\dagger b}_B
\sqrt{\frac{[B]}{B}}= \left[B\right], 
\end{equation}
\begin{equation} \label {eq:q22} 
a a^\dagger = b \sqrt{\frac{[B]}{B}} \sqrt{\frac{[B]}{B}} b^\dagger
=\left[B+1\right]. 
\end{equation}
In the derivation of the last equation the fact that 
for an arbitrary function $f(B)$ one has
\begin{equation}\label{eq:q23}
b^\dagger f(B) =f(B-1)  b^\dagger,  \qquad \mbox{and} \qquad b f(B) =
f(B+1)  b ,
\end{equation}
as well as Eq. (\ref{eq:q15}) have been taken into account.

From Eqs. (\ref{eq:q21}) and (\ref{eq:q22}) one can easily verify that 
\begin{equation}\label{eq:q24}
a a^\dagger -q a^\dagger a = q^{-B}, \qquad a a^\dagger - q^{-1} 
a^\dagger a = q^B,
\end{equation} 
which coincide with Eq. (\ref{eq:q1}). Furthermore using Eqs. (\ref{eq:q14}) 
and (\ref{eq:q20}) one can verify that 
\begin{equation}\label{eq:q25}
[B, a^\dagger]= a^\dagger, \qquad [B, a]=-a, 
\end{equation}
which coincide with Eq. (\ref{eq:q2}). 
Therefore the operators of Eq. (\ref{eq:q20}) can be considered as a 
{\sl representation} of the $q$-deformed boson operators in the 
usual (non-deformed) space 
$\vert n \rangle_c$ of Eq. (\ref{eq:q17}). The operators of Eq. (\ref{eq:q20})
are referred to as {\sl $q$-deforming boson functionals}.
From Eqs. (\ref{eq:q1}) and (\ref{eq:q24}) it is also clear that in this 
representation the operator $N$ is represented by the operator $B$. 

One can now build up the $q$-deformed oscillator states $\vert n \rangle_q$ in
terms of $q$-deforming boson functionals. Starting from Eq. (\ref{eq:q9}) 
and using Eq. (\ref{eq:q20}) one has 
\begin{equation}\label{eq:q26}
\vert n \rangle_q = \frac{(a^\dagger)^n}{\sqrt{[n]!}} \vert 0 \rangle_q 
= \frac{1}{\sqrt{[n]!}}
\underbrace{\sqrt{\frac{[B]}{B}} b^\dagger \sqrt{\frac{[B]}{B}} b^\dagger 
\ldots \sqrt{\frac{[B]}{B}} b^\dagger}_{n \quad {\rm terms}} \vert 0\rangle_c. 
\end{equation}
Making use of the first relation in Eq. (\ref{eq:q23}),  all
 $b^\dagger$ operators can be ``shifted'' to the right, giving 
$$ \vert n \rangle_q = 
\frac{1}{\sqrt{[n]!}} \sqrt{[B] [B-1] [B-2] \ldots [B-n+1] \over 
B (B-1) (B-2) \ldots (B-n+1)} (b^\dagger)^n \vert 0 \rangle_c $$
$$= {1\over \sqrt{[n]!}} \sqrt{ [B] [B-1] [B-2] \ldots [B-n+1] \over 
B (B-1) (B-2) \ldots (B-n+1) } \sqrt{n!} \vert n\rangle_c $$
\begin{equation}\label{eq:q27}
= {1\over \sqrt{[n]!}} \sqrt{ [n] [n-1] [n-2] \ldots [1] \over 
n (n-1) (n-2) \ldots 1} \sqrt{n!} \vert n \rangle_c  = 
\vert n \rangle_c , 
\end{equation}
where in the middle step Eq. (\ref{eq:q17}) has been used. 
We have therefore found that {\it in this representation} the $q$-deformed
oscillator states coincide with the standard oscillator states, a quite 
remarkable result which will play an important role in what follows.  

\section{The Quantum Algebra su$_q$(2)} 

The quantum algebra su$_q$(2) \cite{KulRes,Skly,Jimbo} 
is a $q$-deformation of the Lie algebra su(2).
It is generated by the operators $L_{+}$, $L_{-}$, $L_{0}$, obeying the 
commutation relations (see \cite{PPNP} and references therein) 
\begin{equation} \label{eq:q28}
[L_0,L_{\pm}]=\pm L_{\pm},
\end{equation} %
\begin{equation} \label{eq:q29}
[L_{+},L_{-}]=[2L_0]=\frac{q^{2L_0}-q^{-2L_0}}{q-q^{-1}},
\end{equation} %
where $q$-numbers and $q$-operators are defined as in Eq. (\ref{eq:q4}). 

If the deformation parameter $q$ is not a root of unity (see the discussion 
preceding Eq. (\ref{eq:q6}))
the finite-dimensional irreducible representation $D^\ell_{(q)}$ of su$_q$(2) 
is determined by the highest weight vector $|\ell,\ell\rangle_q$ with 
\begin{equation} \label{eq:q30}
L_{+}|\ell,\ell\rangle_q=0,
\end{equation}  %
and the basis states $|\ell,m\rangle_q$ are expressed as 
\begin{equation} \label{eq:q31}
|\ell,m\rangle_q=\sqrt{\frac{[\ell+m]!}{[2\ell]![\ell-m]!}}
\,(L_{-})^{\ell-m}\,|\ell,\ell\rangle_q. 
\end{equation} %
Then the
explicit form of the irreducible representation (irrep) $D^\ell_{(q)}$ of the 
su$_q$(2) algebra is determined by the equations
\begin{equation} \label{eq:q32}
L_{\pm}|\ell,m\rangle_q=\sqrt{[\ell\mp m][\ell\pm m+1]}\,|\ell,m\pm1\rangle_q, 
\end{equation} %
\begin{equation} \label{eq:q33}
L_0|\ell,m\rangle_q=m\,|\ell,m\rangle_q, 
\end{equation} %
and the dimension of the corresponding representation is the same as in the
non-deformed case, i.e.  ${\rm dim} D^\ell_{(q)}=2\ell+1$ for 
$\ell=0,\frac{1}{2},1,\frac{3}{2},2\ldots$

The second-order Casimir operator of su$_q$(2) is 
$$ C_2^{(q)}=\frac{1}{2}(L_{+}L_{-}+L_{-}L_{+}+[2][L_{0}]^2) $$
\begin{equation} \label{eq:q34}
=L_{-}L_{+}+[L_0][L_0+1]=L_{+}L_{-}+[L_0][L_0-1],
\end{equation} %
while its eigenvalues in the space of the irreducible 
representation $D^\ell_{(q)}$ are $[\ell][\ell+1]$
\begin{equation} \label{eq:q35}
C_2^{(q)} | \ell,m\rangle_q = [\ell][\ell+1] |\ell,m\rangle_q.
\end{equation}

It has been suggested 
that rotational spectra of diatomic molecules \cite{CPL175} and 
deformed nuclei (see \cite{PPNP,RRS} and references therein)  can be 
described by a phenomenological Hamiltonian
based on the symmetry of the quantum algebra su$_q$(2)
\begin{equation} \label{eq:q36}
H=\frac{\hbar^2}{2J_0}\,C_{2}^{(q)}+E_0, 
\end{equation} %
where $C_{2}^{(q)}$ is the second order Casimir operator of Eq. 
(\ref{eq:q34}),
$J_0$ is the moment of inertia for the non-deformed case $q\to1$, and $E_0$ 
is the bandhead energy for a given band.

The eigenvalues of the Hamiltonian of Eq. (\ref{eq:q36}) 
in the basis of Eq. (\ref{eq:q31}) are then
\begin{equation} \label{eq:q37}
E_{\ell}^{(\tau)} = A [\ell] [\ell+1] +E_0,
\end{equation} %
where the definition 
\begin{equation} \label{eq:q38}
A={\hbar^2 \over 2 J_0}
\end{equation}  %
has been used for brevity. 

In the case with $q=e ^\tau,\ \tau\in{\bf R}$ the spectrum of the model 
Hamiltonian of Eq. (\ref{eq:q36})  takes the form
\begin{equation} \label{eq:q39}
E_\ell^{(\tau)}=A \,
\frac{\sinh(\ell\tau)\sinh((\ell+1)\tau)}{\sinh^2(\tau)}+E_0, 
\qquad q=e^{\tau},  
\end{equation} %
while in the case with $q=e^{i\tau}$, $\tau\in{\bf R}$ and 
$q^n\neq 1$, $n\in{\bf N}$ 
the spectrum of the model Hamiltonian of Eq. (\ref{eq:q36})  
takes the form 
\begin{equation} \label{eq:q40}
E_\ell^{(\tau)}=A \,
\frac{\sin(\ell\tau)\sin((\ell+1)\tau)}{\sin^2(\tau)}+E_0, 
\qquad q=e^{i\tau}. 
\end{equation} %

It is known (see \cite{PPNP,CPL175,RRS} and references therein) that 
only the spectrum of Eq. (\ref{eq:q40}) exhibits behavior that is in
agreement with experimentally observed rotational bands.

\section{Representation of su$_q$(2) in Terms of $q$-Deformed 
Boson Operators} 

It has been found \cite{Bie,Mac} that the su$_q$(2) algebra of the last 
section can be represented in terms of two $q$-deformed boson operators
like the ones introduced in Sec. 2. Indeed, one can consider as ``building 
blocks'' the independent $q$-deformed operators 
$a_i^\dagger$ and $a_i$, with $i=1$, 2, which satisfy the commutation 
relations 
\begin{equation}\label{eq:q41} 
a_i a_i^\dagger -q^{\pm 1} a_i^\dagger a_i = q^{\mp N_i}, 
\end{equation}
where $N_i$, $i=1$, 2 are the corresponding $q$-deformed number operators, 
satisfying the commutation relations 
\begin{equation}\label{eq:q42}
[N_i, a_i^\dagger] = a_i^\dagger, \qquad [N_i, a_i] =-a_i.
\end{equation}
One can then verify that the operators 
\begin{equation}\label{eq:q43}
L_+= a_1^\dagger a_2, \qquad L_-=a_1 a_2^\dagger, \qquad L_0={1\over 2}
(N_1-N_2), 
\end{equation}
do satisfy the commutation relations of Eqs. (\ref{eq:q28}) and 
(\ref{eq:q29}). In other words, Eq. (\ref{eq:q43}) expresses a representation 
of the generators of su$_q$(2) in terms of $q$-deformed bosons. 

In this representation the basis states are expressed as 
\begin{equation}\label{eq:q44}
\vert n_1\rangle_q \vert n_2 \rangle_q = {(a_1^\dagger)^{n_1} \over 
\sqrt{[n_1]!}} {(a_2^\dagger)^{n_2} \over \sqrt{[n_2]!}} \vert 0\rangle_q.
\end{equation}
With the identification 
\begin{equation}\label{eq:q45}
n_1= \ell +m, \qquad n_2=\ell -m, 
\end{equation} 
the basis states are put into the form
\begin{equation}\label{eq:q46} 
\vert \ell, m\rangle_q = { (a_1^\dagger)^{\ell+m} \over \sqrt{[\ell+m]!}}
{(a_2^\dagger)^{\ell-m}\over \sqrt{[\ell-m]!}} \vert 0\rangle_q , 
\end{equation}
while the highest weight vector is 
\begin{equation}\label{eq:q47} 
\vert \ell, \ell\rangle_q = {(a_1^\dagger)^{2\ell}\over \sqrt{[2\ell]!}}
\vert 0\rangle_q .
\end{equation}
One can verify that the action of the operators of Eq. (\ref{eq:q43}) 
on the states of Eq. (\ref{eq:q46}) is described by Eqs. (\ref{eq:q32})
and (\ref{eq:q33}). 

\section{Connection of su$_q$(2) to the Lie Algebra su(2)} 

In this section we are going to use both the usual quantum mechanical 
operators of angular momentum, related to the Lie algebra su(2) and 
denoted by $l_+$, $l_-$, $l_0$, 
and the $q$-deformed ones, which are related to su$_q$(2) and denoted by 
$L_+$, $L_-$, $L_0$, as in Sections 4 and 5. 
For brevity we are going 
to call the operators $l_+$, $l_-$, $l_0$ {\sl ``classical''}, 
while the operators $L_+$, $L_-$, $L_0$ will be called 
{\sl ``quantum''}. For the ``classical''
basis the symbol $|l,{\rm m}\rangle _c$ will be used, while the 
``quantum'' basis will be denoted by $|\ell, m\rangle _q$, as in Sections 
4 and 5. 
Therefore $l$ and ${\rm m}$
are the quantum numbers related to the usual quantum mechanical angular 
momentum, which is characterized by the su(2) symmetry, 
while $\ell$ and $ m$ are the quantum numbers related 
to the deformed angular momentum, which is characterized by the su$_q$(2) 
symmetry. 

The ``classical'' operators satisfy the usual su(2) commutation relations
\begin{equation} \label{eq:q48}
[ l_0, l_{\pm}]= \pm  l_{\pm}, \qquad [ l_+,  l_-]=2 l_0,
\end{equation}
while the finite-dimensional irreducible representation $D^l$ of su(2) 
is determined by the highest weight vector $| l, l\rangle _c $ with
\begin{equation} \label{eq:q49}
 l_+|l,l\rangle _c = 0, 
\end{equation} 
and the basis states $|l,{\rm m}\rangle _c$ are expressed as 
\begin{equation} \label{eq:q50}
|l,{\rm m}\rangle _c = \sqrt{ (l+{\rm m})! \over (2l)! (l-{\rm m})!} 
(l_-)^{l-{\rm m}} |l,l\rangle _c. 
\end{equation}
The action of the generators of su(2) on the vectors of the ``classical'' 
basis is described by
\begin{equation} \label{eq:q51}
l_{\pm} | l, {\rm m}\rangle _c = \sqrt{(l\mp {\rm m})(l\pm {\rm m}+1)} 
|l, {\rm m}\pm 1\rangle _c, 
\end{equation} 
\begin{equation} \label{eq:q52}
l_0 |l,{\rm m}\rangle _c = {\rm m} |l,{\rm m}\rangle _c,
\end{equation}
the dimension of the corresponding representation being dim$D^l=2l+1$ 
for $l=0$, ${1\over 2}$, 1, ${3\over 2}$, 2, \dots

The second order Casimir operator of su(2) is 
\begin{equation} \label{eq:q53} 
C_2 = {1\over 2} (l_+ l_-+ l_- l_+) +  l_0^2 = 
l_- l_+ + l_0( l_0+1) 
= l_+  l_- + l_0( l_0-1),  
\end{equation}
where the symbol $1$ is used for the unit operator, 
while its eigenvalues in the space of the irreducible representation $D^l$ 
are $l(l+1)$
\begin{equation} \label{eq:q54} 
 C_2|l, {\rm m}\rangle _c = l(l+1) |l,{\rm m}\rangle _c.
\end{equation}

One can build a representation of su(2) in terms of independent 
boson operators, for which the symbols $b_i^\dagger$ and $b_i$, 
with $i=1,2$ will be used,  
in a way similar to the one used in Sec. 3 for the harmonic oscillator. 
Indeed in the limit $q\to 1$ Eq. (\ref{eq:q41}) is reduced into the usual 
commutation relation for independent boson operators 
\begin{equation} \label{eq:q55}
[b_i, b_i^\dagger] =1, \qquad i=1,2, 
\end{equation}
while Eq. (\ref{eq:q42}) takes the form 
\begin{equation}\label{eq:q56}
[B_i, b_i^\dagger]=b_i^\dagger, \qquad [B_i, b_i]=-b_i, \qquad i=1,2, 
\end{equation}
where by $B_i$ we denote the number operators, for which the following 
relations are valid 
\begin{equation}\label{eq:q57} 
b_i^\dagger b_i =B_i,\qquad  b_i b_i^\dagger =B_i+1,  \qquad i=1,2. 
\end{equation}
Furthermore in the limit $q\to 1$ Eq. (\ref{eq:q43}) is reduced into 
\begin{equation}\label{eq:q58} 
l_+= b_1^\dagger b_2, \qquad l_-=b_1 b_2^\dagger, \qquad 
l_0={1\over 2} (B_1-B_2), 
\end{equation}
which is the well known Schwinger realization of su(2). 
In other words, a representation of the generators of su(2) in terms of boson 
operators is obtained. One can verify that the operators given in 
Eq. (\ref{eq:q58}) do satisfy the commutation relations of Eq. (\ref{eq:q48}).

In this representation the basis states are expressed as 
\begin{equation}\label{eq:q59} 
\vert n_1 \rangle_c \vert n_2 \rangle_c = {(b_1^\dagger)^{n_1} \over 
\sqrt{n_1!}} {(b_2^\dagger)^{n_2} \over \sqrt{n_2!}} \vert 0\rangle_c, 
\end{equation}
which can be viewed as the $q\to 1$ limit of Eq. (\ref{eq:q44}). 
With the identification 
\begin{equation}\label{eq:q60} 
n_1= l+ {\rm m}, \qquad n_2 = l- {\rm m} , 
\end{equation}
the basis states are put into the form 
\begin{equation}\label{eq:q61}
\vert l, {\rm m}\rangle_c = { (b_1^\dagger)^{l+{\rm m}} \over 
\sqrt{(l+{\rm m})!}} {(b_2^\dagger)^{l-{\rm m}} \over 
\sqrt{(l-{\rm m})!}} \vert 0\rangle_c , 
\end{equation}
which can be seen as the $q\to 1$ limit of Eq. (\ref{eq:q46}). 

One can easily verify that the operators of Eq. (\ref{eq:q58}) act on the 
states of Eq. (\ref{eq:q61}) in the way described by Eqs. (\ref{eq:q51})
and (\ref{eq:q52}). 

One can now consider the connection between the ``quantum'' basis of Eq. 
(\ref{eq:q44}) and the ``classical'' basis of Eq. (\ref{eq:q59}). 
Using Eq. (\ref{eq:q27}) one has 
\begin{equation}\label{eq:q62}
\vert n_1\rangle_q \vert n_2 \rangle_q = {(a_1^\dagger)^{n_1} \over 
\sqrt{[n_1]!}} {(a_2^\dagger)^{n_2} \over \sqrt{[n_2]!}} \vert 0\rangle_q 
= {(b_1^\dagger)^{n_1} \over \sqrt{n_1!}} {(b_2^\dagger)^{n_2} \over 
\sqrt{n_2!}} \vert 0\rangle_c = \vert n_1\rangle _c \vert n_2 \rangle_c,
\end{equation}
or, in more compact form, 
\begin{equation}\label{eq:q63} 
\vert n_1 \rangle_q \vert n_2 \rangle_q = \vert n_1 \rangle _c \vert n_2 
\rangle _c .
\end{equation}
In other words, {\it in this boson representation} the ``quantum'' 
states of su$_q$(2) coincide with the ``classical'' states of su(2). 
A direct consequence of this result is obtained by solving Eqs. 
(\ref{eq:q45}) for $\ell$ and $m$
\begin{equation}\label{eq:q64}
\ell= {n_1+n_2 \over 2}, \qquad m = {n_1-n_2\over 2}, 
\end{equation}
and (\ref{eq:q60}) for $l$ and ${\rm m}$
\begin{equation}\label{eq:q65}
l={n_1+n_2 \over 2}, \qquad {\rm m}={n_1-n_2 \over 2}. 
\end{equation}
The fact that the ``quantum'' states and the ``classical'' states 
coincide, as seen in Eq. (\ref{eq:q63}), means that from Eqs. 
(\ref{eq:q64}) and (\ref{eq:q65}) one obtains 
\begin{equation}\label{eq:q66}
\ell = l ={n_1+n_2\over 2}, \qquad m ={\rm m}= {n_1-n_2\over 2}.
\end{equation} 
In other words, the quantum numbers $\ell$ and $m$, appearing in the 
``quantum'' states, are identical with the quantum numbers of the 
physical angular momentum, $l$ and ${\rm m}$, appearing in the 
``classical'' states. It should be remembered that these conclusions 
are valid in the case of $q$ being not a root of unity, as already 
mentioned in Secs. 2 and 4. 

In view of these remarks and using the identifications of Eqs. (\ref{eq:q45})
and (\ref{eq:q60}) one can put Eq. (\ref{eq:q63}) into the form 
\begin{equation}\label{eq:q67p}
\vert \ell, m\rangle_q = { (a_1^\dagger)^{\ell+m}\over \sqrt{[\ell+m]!}}
{(a_2^\dagger)^{\ell-m} \over \sqrt{[\ell-m]!}} \vert 0\rangle_q =
{(b_1^\dagger)^{l+{\rm m}}\over \sqrt{(l+{\rm m})!}} 
{(b_2^\dagger)^{l-{\rm m}}\over \sqrt{(l-{\rm m})!}}=
\vert l, {\rm m}\rangle_c , 
\end{equation}
or, in more compact form, 
\begin{equation}\label{eq:q67} 
\vert \ell, m \rangle_q = \vert l, {\rm m}\rangle_c . 
\end{equation}

The action of the generators and of the second-order Casimir operator 
of su$_q$(2) on the ``classical'' states 
can be seen using Eqs. (\ref{eq:q32}), (\ref{eq:q33}), (\ref{eq:q35}), 
(\ref{eq:q66}), and (\ref{eq:q67})
\begin{equation}\label{eq:q68}
L_{\pm} \vert l, {\rm m}\rangle_c = L_{\pm } \vert \ell, m\rangle_q =
\sqrt{[\ell\mp m][\ell\pm m+1]} \vert \ell, m\pm 1\rangle _q =
\sqrt{[l \mp {\rm m}][l\pm {\rm m}+1]} \vert l, {\rm m}\pm 1\rangle_c, 
\end{equation}
or, in short, 
\begin{equation}\label{eq:q69}
L_{\pm} \vert l, {\rm m}\rangle_c = \sqrt{[l \mp {\rm m}] [l\pm {\rm m}+1] }
\vert l, {\rm m}\pm 1\rangle_c , 
\end{equation}
and in a completely analogous way
\begin{equation}\label{eq:q70}
L_0 \vert l, {\rm m}\rangle_c = {\rm m} \vert l, {\rm m}\rangle_c, 
\end{equation}
\begin{equation}\label{eq:q71} 
C_2^{(q)} \vert l, {\rm m}\rangle_c = [l][l+1] \vert l, {\rm m}\rangle_c . 
\end{equation}

In a similar way one can obtain from Eqs. (\ref{eq:q51}), (\ref{eq:q52}), 
(\ref{eq:q54}), (\ref{eq:q66}), 
and (\ref{eq:q67}) the action of the generators and of the 
second-order Casimir operator of su(2) on the ``quantum'' states 
\begin{equation}\label{eq:q72}
l_{\pm} \vert \ell, m\rangle_q = \sqrt{(\ell\mp m)(\ell\pm m+1)} 
\vert \ell, m\pm 1\rangle_q , 
\end{equation}
\begin{equation}\label{eq:q73}
l_0 \vert \ell, m\rangle_q = m \vert \ell, m\rangle_q , 
\end{equation}
\begin{equation}\label{eq:q74} 
C_2 \vert \ell, m\rangle_q = \ell(\ell+1) \vert \ell, m\rangle_q. 
\end{equation}

As a by-product, one can now show that the operators $\hat L_+$ and $\hat l_+$ 
do not commute 
$$[\hat L_+, \hat l_+]|l, {\rm m}\rangle_c = \hat L_+ \hat l_+ 
|l, {\rm m}\rangle_c - \hat l_+ \hat L_+ | l, {\rm m}\rangle _c$$
$$= \hat L_+ \sqrt{ (l-{\rm m})(l+{\rm m}+1)} |l, {\rm m}+1\rangle_c 
-\hat l_+ \sqrt{[l-{\rm m}][l+{\rm m}+1]} |l, {\rm m}+1\rangle_c $$
$$= (\sqrt{[l-{\rm m}-1][l+{\rm m}+2]} \sqrt{(l-{\rm m})(l+{\rm m}+1)} $$
\begin{equation} \label{eq:q75} 
-\sqrt{(l-{\rm m}-1)(l+{\rm m}+2)} \sqrt{[l-{\rm m}][l+{\rm m}+1]})
|l, {\rm m}+2\rangle _c \neq 0.
\end{equation}
In the same way one can see that 
\begin{equation} \label{eq:q76}
[\hat L_-, \hat l_-]|l, {\rm m}\rangle _c \neq 0. 
\end{equation}
It should be mentioned at this point that the direct connection between 
the generators of su$_q$(2) and the generators of su(2) has been given 
in terms of $q$-deforming functionals 
in Refs. \cite{CGZ,CZPLB}, without any use of boson representations. 

\section{Rotational Invariance of the su$_q$(2) Hamiltonian} 

Using Eqs. (\ref{eq:q69}), (\ref{eq:q71}) 
one can verify that the operator $ C_2^{(q)}$ commutes with 
the generators $L_+$, $L_-$, $L_0$ of su$_q$(2), 
i.e. that $C_2^{(q)}$ is the second 
order Casimir operator of su$_q$(2). Indeed one has 
$$[C_2^{(q)}, L_+]|l, {\rm m}\rangle_c = C_2^{(q)} L_+|l, 
{\rm m}\rangle_c -  L_+ C_2^{(q)} |l, {\rm m}\rangle_c $$
$$= C_2^{(q)} \sqrt{[l-{\rm m}][l+{\rm m}+1]} |l, {\rm m}+1\rangle _c -
L_+[l][l+1] |l, {\rm m}\rangle _c $$
\begin{equation} \label{eq:q77}
=[l][l+1] \sqrt{[l-{\rm m}][l+{\rm m}+1]} |l, {\rm m}+1\rangle _c 
-\sqrt{[l-{\rm m}][l+{\rm m}+1]} [l][l+1] |l, {\rm m}+1\rangle _c =0.
\end{equation}
In exactly the same way one can prove that 
\begin{equation} \label{eq:q78}
[C_2^{(q)}, \hat L_-] |l, {\rm m}\rangle_c =0, 
\end{equation}
while in addition, using Eqs. (\ref{eq:q70}) and (\ref{eq:q71}), one has 
$$[C_2^{(q)}, L_0] \vert l, {\rm m}\rangle_c = C_2^{(q)}  L_0 
\vert l, {\rm m}\rangle_c - L_0  C_2^{(q)} \vert l, {\rm m}\rangle _c $$
\begin{equation} \label{eq:q79}
= [l][l+1] {\rm m} \vert l, {\rm m}\rangle_c - m [l][l+1] \vert 
l, {\rm m}\rangle_c =0. 
\end{equation}

Thus we have verified that the operator $C_2^{(q)}$ is the second order 
Casimir operator of su$_q$(2). We are now going to prove that the operator 
$ C_2^{(q)}$ 
commutes also with the generators $ l_+$, $ l_-$, $ l_0$ 
of the usual su(2) algebra. Indeed using Eqs. (\ref{eq:q51}) 
and (\ref{eq:q71})   one has 
$$[C_2^{(q)}, l_+] |l, {\rm m}\rangle_c = C_2^{(q)} l_+ 
|l, {\rm m}\rangle_c - l_+ C_2^{(q)} |l, {\rm m}\rangle _c $$
$$= C_2^{(q)} \sqrt{(l-{\rm m})(l+{\rm m}+1)} |l, {\rm m}+1\rangle_c 
- l_+[l][l+1] |l, {\rm m}\rangle _c $$
\begin{equation} \label{eq:q80}
= [l][l+1] \sqrt{(l-{\rm m})(l+{\rm m}+1)} |l, {\rm m}+1\rangle _c 
-\sqrt{(l-{\rm m})(l+{\rm m}+1)} [l][l+1] |l, {\rm m}+1\rangle _c =0.
\end{equation}
In exactly the same way one can prove that 
\begin{equation} \label{eq:q81} 
[ C_2^{(q)}, l_-] |l, {\rm m}\rangle _c =0, 
\end{equation}
while the relation
\begin{equation} \label{eq:q82} 
[C_2^{(q)},  l_0] |l, {\rm m}\rangle _c =0 
\end{equation}
occurs from Eqs. (\ref{eq:q52}) and (\ref{eq:q71}).

The following comments are now in place: 
 
a) The fact that the operator $C_2^{(q)}$ 
commutes with the generators of su(2) implies that 
this operator is a function of the second order Casimir operator of 
su(2), given in Eq. (\ref{eq:q53}). 
As a consequence, it should be possible to express
the eigenvalues of $ C_2^{(q)}$, which are $[l][l+1]$  (as we have seen in 
Eq. (\ref{eq:q71})~), in terms of the eigenvalues of $ C_2$, which 
are $l(l+1)$ (as we have seen in Eq. (\ref{eq:q54})~). 
This task will be undertaken in the next section. 

b) Eqs. (\ref{eq:q80})-(\ref{eq:q82}) 
also tell us that the Hamiltonian of Eq. (\ref{eq:q36}) 
commutes with the generators of the usual su(2) algebra, i.e. it is 
rotationally invariant. The Hamiltonian of Eq. (\ref{eq:q36}) does not 
break rotational symmetry. It corresponds to a function of the second order 
Casimir operator of the usual su(2) algebra. This function, however, 
has been chosen in an appropriate way, in order to guarantee that 
the Hamiltonian of Eq. (\ref{eq:q36}) is also invariant under a more
complicated symmetry, namely the symmetry su$_q$(2). 

\section{Exact Expansion of the su$_q$(2) Spectrum}

Let us consider the spectrum of Eq. (\ref{eq:q40}), which has been found 
relevant 
to rotational molecular and nuclear spectra, assuming for simplicity $E_0=0$
and $\tau > 0$. Since the Hamiltonian of Eq. (\ref{eq:q36}) is invariant 
under su(2), as we have seen in the previous section, it should be possible 
in principle to express it as a function of the Casimir operator $C_2$ 
of the usual su(2) algebra. As a consequence, it should also be possible 
to express the eigenvalues of this Hamiltonian, given in Eq. (\ref{eq:q40}),
as a function of the eigenvalues of the Casimir operator of the usual 
su(2), i.e. as a function of $\ell(\ell+1)$.    
This is at first sight 
a nontrivial task, since in Eq. (\ref{eq:q40}) two different functions 
of the variable $\ell$ appear, while we are in need of a single function 
of the variable $\ell(\ell+1)$, which is related to the length of the 
angular momentum vector.  
In order to represent the expression of Eq. (\ref{eq:q40}) 
as a power series of the variable $\ell(\ell+1)$, one can use the identity
\begin{equation} \label{eq:q83}
\sin(\ell\tau)\sin((\ell+1)\tau) 
=\frac{1}{2}\left\{\cos(\tau)-\cos((2\ell+1)\tau)\right\}.
\end{equation} %
The last expression is an even function of $(2\ell+1)$. Therefore it is 
possible to express it as a power series in $(2\ell+1)^2 = 4 \ell (\ell+1)+1$.
It turns out that the coefficients of the relevant expansion 
in powers of $\ell(\ell+1)$ can be expressed 
in terms of the spherical Bessel functions of  the first kind $j_n(x)$ 
\cite{AbrSte}, which are determined through the generating function 
\begin{equation} \label{eq:q84}
{1\over x} \cos\sqrt{x^2-2xt} = \sum_{n=0}^\infty j_{n-1}(x) {t^n \over n!},
\end{equation} %
and are characterized by the asymptotic behavior
\begin{equation} \label{eq:q85} 
j_n(x)\approx\frac{x^n}{(2n+1)!!}, \qquad x\ll 1 .
\end{equation}
Performing the substitutions
\begin{equation} \label{eq:q86} 
x=\tau, \qquad t=-2\tau \ell (\ell+1) ,
\end{equation}  
which imply 
\begin{equation} \label{eq:q87}  
x^2-2xt=\tau^2(2\ell+1)^2,  
\end{equation}
one gets the expression
\begin{equation} \label{eq:q88} 
\frac{1}{\tau}\cos((2\ell+1)\tau)=
\sum_{n=0}^{\infty}\frac{(-2\tau)^n}{n!}\,j_{n-1}(\tau)\,\{\ell(\ell+1)\}^n, 
\end{equation} %
which in the special case of $\ell=0$ reads
\begin{equation} \label{eq:q89} 
{1\over \tau} \cos\tau = j_{-1}(\tau),
\end{equation}
in agreement with the definition  \cite{AbrSte}
\begin{equation} \label{eq:q90} 
j_{-1}(x) = {\cos x \over x}.
\end{equation}
Substituting Eqs. (\ref{eq:q88}) and (\ref{eq:q89}) in Eq. (\ref{eq:q83}),
and taking into account that \cite{AbrSte}
\begin{equation} \label{eq:q91}
j_0(x) = {\sin x \over x},
\end{equation} %
Eq. (\ref{eq:q40}) takes the form 
\begin{equation} \label{eq:q92} 
E_{\ell}^{(\tau)}=\frac{A}{j_0^2(\tau)}\sum_{n=0}^{\infty}
\frac{(-1)^n(2\tau)^n}{(n+1)!}\,j_n(\tau)\,\{\ell(\ell+1)\}^{n+1}, 
\end{equation} %
which is indeed an expansion in terms of $\ell(\ell+1)$. 

\section{Approximate Expansion of the su$_q$(2) Spectrum} 

We are now going to consider an approximate form of this expansion, which 
will allow us to connect the present approach to the description of 
rotational spectra proposed by Amal'sky \cite{Amal}. 

For ``small deformation'', i.e. for $\tau\ll1$, one can use the asymptotic 
expression of Eq. (\ref{eq:q85}). 
Keeping only the terms of the lowest order one then 
obtains the following approximate series
\begin{equation} \label{eq:q93}
E_{\ell}^{(\tau)}\approx A\,\sum_{n=0}^{\infty}
\frac{(-1)^n(2\tau)^{2n}}{(n+1)(2n+1)!}\,\{\ell(\ell+1)\}^{n+1}, 
\end{equation} %
where use of the identity 
\begin{equation} \label{eq:q94}
2^n (n+1)! (2n+1)!! = (n+1) (2n+1)!
\end{equation} %
has been made. 
The first few terms of this expansion are
\begin{equation} \label{eq:q95} 
E_{\ell}^{(\tau)}\approx A\Bigl(\,\ell(\ell+1)
-\frac{\tau^2}{3}\{\ell(\ell+1)\}^2 +\frac{2\tau^4}{45}\{\ell(\ell+1)\}^3
-\frac{\tau^6}{315}\{\ell(\ell+1)\}^4+\ldots\Bigr),
\end{equation} %
in agreement with the findings of Ref. \cite{PLB251}. 

One can now observe that the expansion appearing in Eq. (\ref{eq:q93})   
is similar to the power series of the function
\begin{equation} \label{eq:q96} 
\sin^2 x=\frac{1}{2}(1-\cos 2x)=
\sum_{k=1}^{\infty}(-1)^{k+1}2^{2k-1}\frac{x^{2k}}{(2k)!}.
\end{equation} %
Then, performing the auxiliary substitution
\begin{equation} \label{eq:q97} 
\xi=\sqrt{\ell(\ell+1)}, \qquad\qquad \eta=\ell(\ell+1)=\xi^2, 
\end{equation} %
one can put the expansion of Eq. (\ref{eq:q93}) in the form 
\begin{equation} \label{eq:q98} 
E_\ell^{(\tau)} \approx A\,\frac{\sin^2(\tau\xi)}{\tau^2}=
\frac{\hbar^2}{2J_0}\frac{\sin^2(\tau\sqrt{\ell(\ell+1)})}{\tau^2},
\qquad q=e^{i\tau}. 
\end{equation}  %
This result is similar to the expression proposed for the unified 
description of nuclear rotational spectra by G. Amal'sky \cite{Amal}
\begin{equation} \label{eq:q99}
E_\ell=\varepsilon_0\,\sin^2\left(\frac{\pi}{N}\sqrt{\ell(\ell+1)}\right), 
\end{equation}
where $\varepsilon_0$ is a phenomenological constant 
($\varepsilon_0\approx 6.664$ MeV) which remains the same for all nuclei, 
while $N$ is a free parameter varying from one nucleus to the other. 

\section{Irreducible Tensor Operators under su$_q$(2)}

A different path towards the construction of a Hamiltonian appropriate 
for the description of rotational spectra can be taken through the 
construction of irreducible tensor operators under su$_q$(2)
\cite{STK593,STK690}. In this discussion we limit ourselves to real values 
of $q$, i.e. to $q=e^\tau$ with $\tau$ being real, as in Refs. 
\cite{STK593,STK690}.  

An irreducible tensor operator of rank $k$ is the 
set of $2k+1$ operators $T^{(q)}_{k,\kappa}$ ($\kappa=k$, $k-1$, $k-2$, 
$\ldots$, $-k$), which satisfy with the generators of the su$_q$(2) 
algebra the commutation relations \cite{STK593,STK690} 
\begin{equation} \label{eq:q100}
[L_0, T^{(q)}_{k,\kappa}]= \kappa T^{(q)}_{k,\kappa},
\end{equation}
\begin{equation} \label{eq:q101}
[L_{\pm}, T^{(q)}_{k,\kappa}]_{q^\kappa} = \sqrt{[k\mp \kappa] [k\pm \kappa+1]}
T^{(q)}_{k,\kappa\pm 1} q^{-L_0},
\end{equation}
where $q$-commutators are defined by  
\begin{equation} \label{eq:q102} 
[A,B]_{q^\alpha} = A B-q^\alpha B A.
\end{equation}
It is clear that in the limit $q\rightarrow 1$ these commutation relations 
reduce to the usual ones, which occur in the definition of irreducible 
tensor operators under su(2). 
It should also be noticed that the operators
\begin{equation} \label{eq:q103} 
R^{(q)}_{k,\kappa} = (-1)^\kappa q^{-\kappa} (T^{(q)}_{k,-\kappa})^\dagger,
\end{equation}
where $\dagger$ denotes Hermitian conjugation, satisfy the same commutation
relations (\ref{eq:q100}), (\ref{eq:q101}) as the operators 
$T^{(q)}_{k,\kappa}$, i.e. the operators $R^{(q)}_{k,\kappa}$ also form 
an irreducible tensor operator of rank $k$ under su$_q$(2). 

We can construct an irreducible tensor operator of rank 1 using as building 
blocks the generators of su$_q$(2). This irreducible tensor operator will
consist of the operators $J_{+1}$, $J_{-1}$, $J_0$, which should satisfy 
the commutation relations 
\begin{equation} \label{eq:q104} 
[L_0, J_m]= m J_m,
\end{equation}
\begin{equation} \label{eq:q105} 
[L_{\pm}, J_m ]_{q^m} = \sqrt{[1\mp m] [2\pm m]} J_{m\pm 1} q^{-L_0},
\end{equation}
which are a special case of Eqs. (\ref{eq:q100}), (\ref{eq:q101}), while 
the relevant Hermitian conjugate operators will be 
\begin{equation} \label{eq:q106} 
(J_m)^\dagger = (-1)^m q^{-m} J_{-m},
\end{equation} 
which is a consequence of Eq. (\ref{eq:q103}). 
It turns out \cite{STK593,STK690,JPA6939}
that the explicit form of the relevant operators is
\begin{equation} \label{eq:q107}
J_{+1}= -{1\over \sqrt{[2]}} q^{-L_0} L_+, 
\end{equation}
\begin{equation} \label{eq:q108}
J_{-1}={1\over \sqrt{[2]}} q^{-L_0} L_-,
\end{equation} 
\begin{equation} \label{eq:q109} 
J_0={1\over [2]} (qL_+ L_- -q^{-1}L_- L_+) 
={1\over [2]} \left( q [2L_0]+(q-q^{-1}) ( C_2^{(q)}-[L_0][L_0+1])\right),
\end{equation}
while the Hermitian conjugate operators are 
\begin{equation} \label{eq:q110}
(J_{+1})^\dagger =-q^{-1} J_{-1}, \quad (J_{-1})^\dagger = -q J_{+1}, \quad
(J_0)^\dagger = J_0. 
\end{equation}
It is clear that in the limit $q\rightarrow 1$ these results reduce to the 
usual expressions for spherical tensors of rank 1 under su(2), formed out 
of the usual angular momentum operators 
\begin{equation} \label{eq:q111}
J_+=-{l_+\over \sqrt{2}}=-{l_x+il_y \over \sqrt{2}}, \qquad 
J_-={l_-\over \sqrt{2}}={l_x-il_y\over \sqrt{2}}, \qquad J_0 = l_0,
\end{equation} 
\begin{equation} \label{eq:q112}
(J_+)^\dagger = -J_-, \quad (J_-)^\dagger = -J_+, \quad (J_0)^\dagger =J_0. 
\end{equation}

The commutation relations among the operators $J_{+1}$, $J_{-1}$, $J_0$ can be 
obtained using Eqs. (\ref{eq:q107})-(\ref{eq:q109}) and (\ref{eq:q104}), 
(\ref{eq:q105}), as well as the fact that from Eq. (\ref{eq:q28}) one has 
\begin{equation} \label{eq:q113}
[L_0,L_+]= L_+\Rightarrow L_0 L_+ = L_+(L_0+1) \Rightarrow f(L_0) L_+
= L_+ f(L_0+1),
\end{equation}
\begin{equation} \label{eq:q114}
[L_0, L_-]= -L_- \Rightarrow L_0 L_- = L_- (L_0-1) \Rightarrow f(L_0) L_-
= L_- f(L_0-1),
\end{equation}
where $f(x)$ is any function which can be written as a Taylor expansion 
in powers of $x$.  Indeed one finds  
\begin{equation} \label{eq:q118} 
[J_{+1},J_0]=-q^{-2L_0+1}J_{+1}, \quad 
[J_{-1},J_0]=q^{-2L_0-1}J_{-1}, \quad 
[J_{+1},J_{-1}]=-q^{-2L_0}J_{0}. 
\end{equation}
In the limit $q\to 1$ these results reduce to the usual commutation relations 
related to spherical tensor operators under su(2)
\begin{equation} \label{eq:q119}
[J_+, J_0]= - J_+, \quad [J_-, J_0]=J_-, \quad [J_+, J_-]=-J_0. 
\end{equation}
It is clear that the commutation relations of Eq. (\ref{eq:q118}) are 
different from these of Eqs. (\ref{eq:q28}), (\ref{eq:q29}), as it is expected
since the commutation relations of Eq. (\ref{eq:q119}) are different from 
the usual commutation relations of su(2), given in Eq. (\ref{eq:q48}).

One can now try to build out of these operators the scalar square of the 
angular momentum operator.  For this purpose one needs the definition 
of the tensor product of two irreducible tensor operators, which has the form
\cite{STK593,STK690,JPA6939,STK1068,STK1599,JPG1931}
\begin{equation} \label{eq:q121}
[ A^{(q)}_{j_1} \otimes B^{(q)}_{j_2} ] ^{(1/q)}_{j,m} = \sum_{m_1,m_2} 
\langle j_1 m_1 j_2 m_2 | j m\rangle_{1/q} A^{(q)}_{j_1,m_1} B^{(q)}_{j_2,m_2}.
\end{equation}
One should observe that the irreducible tensor operators $A^{(q)}_{j_1}$ 
and $B^{(q)}_{j_2}$, which correspond to the deformation parameter $q$, 
 are combined into a new irreducible tensor operator
$[A^{(q)}_{j_1} \times B^{(q)}_{j_2}]^{(1/q)}_{j,m}$, 
which corresponds to the deformation parameter $1/q$,  through the use 
of the deformed Clebsch--Gordan coefficients 
$\langle j_1 m_1 j_2 m_2 | j m\rangle _{1/q}$,
which also correspond to the deformation parameter $1/q$. 

Analytic expressions for several $q$-deformed Clebsch--Gordan coefficients, 
as well as their symmetry proporties,  can be found in 
Refs. \cite{STK593,STK1068}. Using the general formulae of Refs. 
\cite{STK593,STK1068} 
we derive here the Clebsch--Gordan coefficients which we will immediately need 
\begin{equation} \label{eq:q122}
\langle 11 10 | 11\rangle _q=q\sqrt{\frac{[2]}{[4]}}, \quad  
\langle 10 11 |11\rangle _q=-q^{-1}\sqrt{\frac{[2]}{[4]}}, 
\end{equation}
\begin{equation} \label{eq:q123}
\langle 10 1-1 |1-1\rangle _q=q\sqrt{\frac{[2]}{[4]}} ,\quad
\langle 1-1 10 |1-1\rangle _q=-q^{-1}\sqrt{\frac{[2]}{[4]}},
\end{equation} 
\begin{equation} \label{eq:q124} 
\langle 11 1-1 |10\rangle _q=\sqrt{\frac{[2]}{[4]}}, \quad
\langle 1-1 11 |10\rangle _q=-\sqrt{\frac{[2]}{[4]}}, \quad 
\langle 10 10 |10\rangle _q=(q-q^{-1})\sqrt{\frac{[2]}{[4]}}.
\end{equation}

Using the definition of Eq. (\ref{eq:q121}), the Clebsch--Gordan coefficients
just given, as well as the commutation relations of Eq. (\ref{eq:q118}), 
one finds the tensor products 
$$ [J\otimes J]_{1,+1}^{(1/q)}=
\langle 11 10 |11\rangle _{1/q} J_{+1}J_{0}+
\langle 10 11 |11\rangle _{1/q} J_{0}J_{+1} $$
\begin{equation} \label{eq:q125}
=-\sqrt{\frac{[2]}{[4]}}
\left\{q^{-2L_0}+(q-q^{-1})J_0\right\}J_{+1}, 
\end{equation}
$$ [J\otimes J]_{1,-1}^{(1/q)}=
\langle 10 1-1 |1-1\rangle _{1/q}J_{0}J_{-1}+
\langle 1-1 10|1-1\rangle _{1/q} J_{-1}J_{0} $$
\begin{equation} \label{eq:q126} 
=-\sqrt{\frac{[2]}{[4]}}
\left\{q^{-2L_0}+(q-q^{-1})J_0\right\}J_{-1}, 
\end{equation}
$$ [J\otimes J]_{1,0}^{(1/q)}=
\langle 11 1-1|10\rangle _{1/q}J_{+1}J_{-1}+
\langle 1-1 11|10\rangle _{1/q}J_{-1}J_{+1}+
\langle 10 10|10\rangle _{1/q}(J_{0})^2 $$
\begin{equation} \label{eq:q127}
=-\sqrt{\frac{[2]}{[4]}}
\left\{q^{-2L_0}+(q-q^{-1})J_0\right\}J_{0}. 
\end{equation}
We remark that all these tensor products are of the general form
\begin{equation} \label{eq:q128} 
[J\otimes J]_{1,m}^{(1/q)}=
-\sqrt{\frac{[2]}{[4]}}\left\{q^{-2L_0}+(q-q^{-1})J_0\right\}J_{m}
= -\sqrt{\frac{[2]}{[4]}} Z J_m, 
\qquad\qquad m=0,\pm1
\end{equation}
where by definition 
\begin{equation} \label{eq:q129}
Z=q^{-2L_0}+(q-q^{-1})J_0.
\end{equation}
One can now prove that the operator $Z$ is a scalar quantity, since it is 
a function of the second order Casimir operator of su$_q$(2), given in 
Eq. (\ref{eq:q34}). Indeed one finds 
\begin{equation} \label{eq:q131} 
Z=q^{-2L_0}+(q-q^{-1})J_0=1+\frac{(q-q^{-1})^2}{[2]}\,C_2 ^{(q)}.
\end{equation}

Since $Z$ is a scalar quantity, symmetric under the exchange 
$q\leftrightarrow q^{-1}$ (as one can see from the last expression 
appearing in the last equation), Eq. (\ref{eq:q128})  can be written 
in the form 
\begin{equation} \label{eq:q132}
\left[ {J\over Z} \otimes {J\over Z} \right]_{1,m}^{(1/q)} = -\sqrt{[2]\over
 [4]} {J_m \over Z} \Rightarrow [J' \otimes J']_{1,m}^{(1/q)} = 
-\sqrt{[2]\over [4]} J'_m,
\end{equation}
where by definition
\begin{equation} \label{eq:q133}
J'_m = {J_m \over Z}, \qquad m=+1,0,-1. 
\end{equation}
It is clear that the operators $J'_m$ also form 
an irreducible tensor operator, since $Z$ is a function of the second order 
Casimir $C_2^{(q)}$ of su$_q$(2), which commutes with the generators $L_+$, 
$L_-$, $L_0$ of su$_q$(2), and therefore does not affect the commutation
relations of Eqs. (\ref{eq:q104}), (\ref{eq:q105}). 

The scalar product of two irreducible tensor operators is defined as 
\cite{STK690,JPG1931} 
\begin{equation} \label{eq:q134}
( A^{(q)}_j \cdot B^{(q)}_j )^{(1/q)} = (-1)^{-j} \sqrt{[2j+1]}
[ A^{(q)}_j \times B^{(q)}_j ]^{(1/q)}_{0,0} 
= \sum_m (-q)^{-m} A^{(q)}_{j,m} B^{(q)}_{j,-m}.
\end{equation}
Substituting the irreducible tensor operators $J_m$ in this definition 
we obtain \cite{STK690}
\begin{equation} \label{eq:q135} 
(J \cdot J)^{(1/q)}= - \sqrt{[3]} [J\times J]_{0,0}^{(1/q)} =
{2\over [2]} C_2^{(q)} + {(q-q^{-1})^2 \over [2]^2} (C_2^{(q)})^2 
= {Z^2 -1 \over (q-q^{-1})^2},  
\end{equation}
where in the last step the identity 
\begin{equation} \label{eq:q136}
Z^2-1 = (Z-1)(Z+1) = {(q-q^{-1})^2 \over [2]} C_2^{(q)} 
\left( 2+{(q-q^{-1})^2 \over [2]} C_2^{(q)} \right),
\end{equation}
has been used, obtained through use of Eq. (\ref{eq:q131}). 
In the same way the irreducible tensor operators $J'_m$ give the result
\begin{equation} \label{eq:q137}
\left( J' \cdot J' \right)^{(1/q)} = {1-Z^{-2} \over (q-q^{-1})^2} . 
\end{equation}
We have therefore determined the scalar square of the angular momentum 
operator. We can assume at this point that this quantity can be used 
(up to an overall constant) as the Hamiltonian for the description of 
rotational spectra, defining 
\begin{equation} \label{eq:q138}
H = A {1-Z^{-2} \over (q-q^{-1})^2},
\end{equation}
where $A$ is a constant, which one can also write in the form
\begin{equation} \label{eq:q139} 
A={\hbar^2 \over 2  J_0},
\end{equation} 
where $ J_0$ is the moment of inertia. 

The eigenvalues $\langle Z\rangle $ of the operator $Z$ 
in the basis $|\ell,m\rangle_q $ can be 
easily found from the last expression given in Eq. (\ref{eq:q131}), 
using the eigenvalues of the Casimir operator $C_2^{(q)}$ in this basis, 
which are $[\ell] [\ell+1]$, as already mentioned in Sec. 4
\begin{equation} \label{eq:q140} 
\langle Z\rangle = 1 + {(q-q^{-1})^2 \over [2]} [\ell] [\ell+1] = 
{1\over [2]}
(q^{2\ell+1}+q^{-2\ell-1}) = {1\over [2]} ([2\ell+2]-[2\ell]).
\end{equation}
The eigenvalues $\langle (J\cdot J)^{(1/q)}\rangle $ of the scalar quantity 
$(J \cdot J)^{(1/q)}$ can be 
found in a similar manner from Eq. (\ref{eq:q135})
\begin{equation} \label{eq:q141}
\langle (J \cdot J)^{(1/q)}\rangle  
= {2\over [2]} [\ell][\ell+1] + {(q-q^{-1})^2 \over [2]^2}
[\ell]^2 [\ell+1]^2 = {[2\ell] [2\ell+2] \over [2]^2} =
[\ell]_{q^2} [\ell+1]_{q^2} ,
\end{equation}
where by definition
\begin{equation} \label{eq:q142} 
[x]_{q^2} = {q^{2x}-q^{-2x}\over q^2 -q^{-2}}.
\end{equation}
Finally, the eigenvalues $\langle H\rangle $ 
of the Hamiltonian can be found by substituting 
the eigenvalues of $Z$ from Eq. (\ref{eq:q140}) into Eq. (\ref{eq:q138})
$$ E = \langle H\rangle = A {1\over (q-q^{-1})^2} \left( 1-{[2]^2 \over 
(q^{2\ell+1}+q^{-2\ell-1})^2 } \right) $$ 
\begin{equation} \label{eq:q143}
=A {1\over 4\sinh^2\tau } \left( 1-{\cosh^2\tau \over \cosh^2((2\ell+1)\tau)}
\right), \qquad q=e^\tau, 
\end{equation}
where in the last step the identities 
\begin{equation} \label{eq:q144} 
q-q^{-1}= 2\sinh\tau, \qquad [2] = q+q^{-1} = 2\cosh\tau,
\end{equation}
\begin{equation} \label{eq:q145} 
q^{2\ell+1}+q^{-2\ell-1} =2\cosh ((2\ell+1)\tau),
\end{equation}
which are valid in the present case of $q=e^\tau$ with $\tau$ being real, 
have been used. 
In the same way one sees that 
\begin{equation} \label{eq:q146}
\langle Z\rangle = {\cosh((2\ell+1)\tau)\over \cosh\tau}.
\end{equation}

The following comments are now in place: 

a) The last expression in Eq. (\ref{eq:q141}) indicates that the eigenvalues 
of the scalar quantity $(J\cdot J)^{(1/q)}$ 
are equivalent to the eigenvalues of the Casimir operator of su$_q$(2) 
(which are $[\ell][\ell+1]$), up to a change 
in the deformation parameter from $q$ to $q^2$. 

b) From Eq. (\ref{eq:q140}) it is clear that the eigenvalues of the 
scalar operator $Z$ go to the limiting value 1 as $q\rightarrow 1$. 
Therefore one can think of $Z$ as a ``unity'' operator. Furthermore the last 
expression in Eq. (\ref{eq:q140}) indicates that 
$\langle Z\rangle $ is behaving like 
a ``measure'' of the unit of angular momentum in the deformed case. 

\section{Rotational Invariance of the su$_q$(2) ITO Hamiltonian} 

In this section the method of Sec. 7 will be used once more. 
We wish to prove that the Hamiltonian of Eq. (\ref{eq:q138}) commutes 
with the generators $l_+$, $l_-$, $l_0$ of the usual su(2) 
algebra, i.e.
with the usual angular momentum operators. 
Taking into account Eq. (\ref{eq:q131}) we see that 
acting on the ``classical'' basis described in Sec. 6 we have
\begin{equation} \label{eq:q147} 
Z |l, {\rm m}\rangle _c = \left( 1+ {(q-q^{-1})^2 \over [2]} 
C_2^{(q)}\right)
| l, {\rm m}\rangle _c = \left(1+{(q-q^{-1})^2 \over [2]}[l] [l+1]\right)
|l, {\rm m}\rangle _c.
\end{equation}
Then using Eq. (\ref{eq:q138}) we see that 
$$ H |l, {\rm m}\rangle _c = {A\over (q-q^{-1})^2} \left(1-{1\over 
Z^2}\right) 
|l, {\rm m}\rangle _c $$
\begin{equation} \label{eq:q148} 
= {A\over (q-q^{-1})^2} \left(1-{1\over 
\left( 1+ {(q-q^{-1})^2 \over [2]} [l] [l+1]\right)^2} \right) 
|l, {\rm m}\rangle _c.
\end{equation}
Using this result, as well as Eq. (\ref{eq:q51}), one finds 
$$[H, l_+] |l, {\rm m}\rangle _c =  H  l_+ |l, 
{\rm m}\rangle _c - l_+  H |l, {\rm m}\rangle _c $$
$$=  H \sqrt{(l-m)(l+m+1)} |l, {\rm m}+1\rangle _c $$
$$- l_+ {A\over (q-q^{-1})^2} 
\left(1 -{1\over \left(1+{(q-q^{-1})^2\over [2]} [l][l+1] \right)^2}\right) 
|l, {\rm m}\rangle _c$$
$$={A\over (q-q^{-1})^2} \left(1-{1\over \left(1+{(q-q^{-1})^2 \over [2]} 
[l][l+1]\right)^2}\right) \sqrt{(l-m)(l+m+1)} |l, {\rm m}+1\rangle _c $$
\begin{equation} \label{eq:q149} 
-\sqrt{(l-m)(l+m+1)} {A\over (q-q^{-1})^2} \left(1-{1\over \left(1+ 
{(q-q^{-1})^2\over [2]} [l][l+1]\right)^2 }\right) |l, {\rm m}+1\rangle_c =0. 
\end{equation}
In exactly the same way, using Eqs. (\ref{eq:q51}), (\ref{eq:q52})  
and (\ref{eq:q148}), one finds that 
\begin{equation} \label{eq:q150} 
[ H, l_-] |l, {\rm m}\rangle_c =0, \qquad [ H,  l_0] 
|l, {\rm m}\rangle_c =0.
\end{equation}
We have thus proved that the Hamiltonian of Eq. (\ref{eq:q138}) is invariant 
under usual angular momentum. This result is expected, since the Hamiltonian 
is a function of the operator $Z$, which in turn (as seen from Eq. 
(\ref{eq:q131})~) is a function of the second order Casimir operator 
of su$_q$(2), $C_2^{(q)}$, which was proved to be rotationally invariant 
in Section 7. 

Since the Hamiltonian of Eq. (\ref{eq:q138}) is rotationally invariant, 
it should be possible to express 
it as a function of $C_2$ (the second order Casimir 
operator of su(2)~). It should also be possible to express the eigenvalues 
of the Hamiltonian of Eq. (\ref{eq:q138}) as a function of $l(l+1)$, i.e. 
as a function of the eigenvalues of $C_2$. This task will be undertaken 
in the following section. 

For completeness we mention that using Eqs. (\ref{eq:q69}), 
(\ref{eq:q70}), and 
(\ref{eq:q148}) one can prove in an analogous way that 
\begin{equation} \label{eq:q151}
[ H, L_+] |l, {\rm m}\rangle_c =0, \qquad 
[ H, L_-] |l, {\rm m}\rangle_c =0, \qquad 
[ H, L_0] |l, {\rm m}\rangle_c =0,
\end{equation}
i.e. that the Hamiltonian of Eq. (\ref{eq:q138}) commutes with the 
generators of su$_q$(2) as well. Then from Eqs. 
(\ref{eq:q107})-(\ref{eq:q109}) it is clear that in addition one has 
\begin{equation} \label{eq:q152} 
[H, J_+] |l, {\rm m}\rangle _c=0, \qquad 
[H, J_-] |l, {\rm m}\rangle_c =0, \qquad  
[H, J_0] |l, {\rm m}\rangle _c =0. 
\end{equation}
Then from Eqs. (\ref{eq:q133}) and (\ref{eq:q147}) one furthermore obtains
\begin{equation} \label{eq:q153}
[H, J'_+]|l, {\rm m}\rangle _c =0, \qquad 
[H, J'_-] |l, {\rm m}\rangle _c =0, \qquad 
[H, J'_0] |l, {\rm m}\rangle _c=0. 
\end{equation}

\section{Exact Expansion of the su$_q$(2) ITO Spectrum}

Since the Hamiltonian of Eq. (\ref{eq:q138}) is invariant under su(2), 
as we have seen in the last section, it should be possible to write 
its eigenvalues (given in Eq. (\ref{eq:q143})~) as an expansion in terms 
of $\ell(\ell+1)$. At this point it is useful to observe that 
in Eq. (\ref{eq:q143}) an even function of the variable $(2\ell+1)$ appears, 
which can therefore be expressed as a power series in 
$(2\ell+1)^2 = 4\ell(\ell+1)+1$.   
In this direction it turns out that one should use 
the Taylor expansion \cite{AbrSte}
\begin{equation}\label{eq:q154}
\tanh x=\sum_{n=1}^{\infty}\frac{2^{2n}(2^{2n}-1)B_{2n}}{(2n)!}
\,x^{2n-1}
=\sum_{n=0}^{\infty}\frac{2^{2n+2}(2^{2n+2}-1)B_{2n+2}}{(2n+2)!}
\,x^{2n+1}
\quad,\qquad |x|<\frac{\pi}{2}, 
\end{equation}
where $B_{n}$ are the Bernoulli numbers \cite{AbrSte}, defined through the
generating function
\begin{equation} \label{eq:q155} 
\frac{x}{e^x-1}=\sum_{n=0}^{\infty}B_n\frac{x^n}{n!}, 
\end{equation}
the first few of them being 
$$ B_0=1,\quad B_1=-\frac{1}{2},\quad B_2=\frac{1}{6},
\quad B_4=-\frac{1}{30},\quad B_6=\frac{1}{42},
\quad B_8=-\frac{1}{30},\quad B_{10}=\frac{5}{66},\ldots, $$
\begin{equation} \label{eq:q156}
B_{2n+1}=0 \quad {\rm for} \quad n=1,2,\ldots
\end{equation}
From Eq. (\ref{eq:q154}) the following identities, concerning the derivatives 
of $\tanh x$, occur 
\begin{equation} \label{eq:q157}
(\tanh x)'=\frac{1}{\cosh ^2x}=1-\tanh ^2x
=\sum_{n=0}^{\infty}
\frac{2^{2n+2}(2^{2n+2}-1)B_{2n+2}}{(2n)!(2n+2)}\,x^{2n}, 
\end{equation}
\begin{equation} \label{eq:q158}
(\tanh x)''= -2{\tanh x \over \cosh^2x} = -2 {\sinh x \over \cosh^2 x} 
=\sum_{n=0}^{\infty}
\frac{2^{2n+4}(2^{2n+4}-1)B_{2n+4}}{(2n+1)!(2n+4)}\,x^{2n+1}.
\end{equation}
From these equations the following auxiliary identities occur 
\begin{equation} \label{eq:q159}
\frac{\sinh x}{x \ \cosh^3x}=-\frac{1}{2x}(\tanh x)''
=\sum_{n=0}^{\infty}
\frac{2^{2n+3}(1-2^{2n+4})B_{2n+4}}{(2n+1)!(2n+4)}\,x^{2n}, 
\end{equation}
\begin{equation} \label{eq:q160}
\tanh^2x=1-\frac{1}{\cosh^2x}
=\sum_{n=0}^{\infty}
\frac{2^{2n+4}(1-2^{2n+4})B_{2n+4}}{(2n+2)!(2n+4)}\,x^{2n+2}. 
\end{equation}
The expression for the energy, given in Eq. (\ref{eq:q143}), can be put 
in the form  
\begin{equation} \label{eq:q161}
{E\over A} =
\left(\frac{{\rm cosh}^2\tau\,.\tau^2}{{\rm sinh}^2\tau}\right)
\frac{1}{(2\tau)^2}\left\{\frac{1}{{\rm cosh}^2\tau}
-\frac{1}{{\rm cosh}^2((2\ell+1)\tau)}\right\}. 
\end{equation}
Denoting
\begin{equation} \label{eq:q162}
z=(2\ell+1)\tau, \qquad x=\ell(\ell+1), 
\end{equation}
which imply 
\begin{equation} \label{eq:q163} 
z^2=(4x+1)\tau^2, \qquad\qquad 
z^{2n}=\tau^{2n}\sum_{k=0}^{n}{n\choose k}\,2^{2k}\,x^k, 
\end{equation}
(the latter through use of the standard binomial formula), 
one obtains from Eq. (\ref{eq:q157}) the expansion 
$$ \frac{1}{\cosh^2((2\ell+1)\tau)}=
\frac{1}{\cosh^2z}=\sum_{n=0}^{\infty}
\frac{2^{2n+2}(2^{2n+2}-1)B_{2n+2}}{(2n)!(2n+2)}\,z^{2n} $$
\begin{equation} \label{eq:q164} 
=\sum_{n=0}^{\infty}\;
\underbrace{\frac{2^{2n+2}(2^{2n+2}-1)B_{2n+2}}{(2n)!(2n+2)}
\,\tau^{2n}}_{a_n}
\;\sum_{k=0}^{n}\;
\underbrace{{n\choose k}\,2^{2k}}_{b_{n,k}}\;x^k
=\sum_{n=0}^{\infty} a_n\, \sum_{k=0}^n \, b_{n,k}\,  x^k. 
\end{equation}
The double sum appearing in the last expression can be rearranged using the 
general procedure 
$$ S=\sum_{n=0}^{\infty}a_n\sum_{k=0}^n b_{n,k}\,x^k $$
$$ = a_0b_{00}+a_1(b_{10}+b_{11}x)+a_2(b_{20}+b_{21}x+b_{22}x^2)
+a_3(b_{30}+b_{31}x+b_{32}x^2+b_{33}x^3)+\ldots $$
$$=(a_0b_{00}+a_1b_{10}+a_2b_{20}+a_3b_{30}+\ldots)
+(a_1b_{11}+a_2b_{21}+a_3b_{31}+a_4b_{41}+\ldots)\,x $$
$$+(a_2b_{22}+a_3b_{32}+a_4b_{42}+a_5b_{52}+\ldots)\,x^2
+(a_3b_{33}+a_4b_{43}+a_5b_{53}+a_6b_{63}+\ldots)\,x^3
\,+\,\ldots $$
\begin{equation}\label{eq:q165}
=\sum_{n=0}^{\infty}\,\underbrace{\left\{
\sum_{k=n}^{\infty} a_{k}\,b_{k,n}\right\}}_{c_n}\,x^n
=\sum_{n=0}^{\infty}c_{n}\,x^n, 
\end{equation}
where
\begin{equation} \label{eq:q166} 
c_{n}=\sum_{k=n}^{\infty} a_{k}\,b_{k,n}
=\sum_{k=0}^{\infty} a_{n+k}\,b_{n+k,n}. 
\end{equation}
Applying this general procedure in the case of Eq. (\ref{eq:q164}) we obtain 
\begin{equation} \label{eq:q167} 
{1\over \cosh^2((2\ell+1)\tau)} = {1\over \cosh^2z} = \sum_{n=0}^\infty 
c_n x^n, 
\end{equation}
where
$$ c_{n}=\sum_{k=0}^{\infty} a_{n+k}\,b_{n+k,n}\nonumber\\[2ex]
=\sum_{k=0}^{\infty}
\frac{2^{2n+2k+2}(2^{2n+2k+2}-1)B_{2n+2k+2}}{(2n+2k)!(2n+2k+2)}
\,\tau^{2n+2k}\,{n+k\choose n}\,2^{2n} $$
\begin{equation} \label{eq:q168} 
=(2\tau)^{2n}\,\sum_{k=0}^{\infty}
\frac{2^{2n+2k+2}(2^{2n+2k+2}-1)B_{2n+2k+2}}{(2n+2k)!(2n+2k+2)}
\,{n+k\choose n}\,\tau^{2k}. 
\end{equation}
The first term in Eq. (\ref{eq:q167}) is
\begin{equation} \label{eq:q169} 
c_0=\sum_{k=0}^{\infty}
\frac{2^{2k+2}(2^{2k+2}-1)B_{2k+2}}{(2k)!(2k+2)}
\,\tau^{2k}=\frac{1}{{\rm cosh}^2\tau}.
\end{equation}
Then one has 
$$ \frac{1}{(2\tau)^2}\left\{\frac{1}{{\rm cosh}^2\tau}
-\frac{1}{{\rm cosh}^2((2\ell+1)\tau)}\right\} 
=-\frac{1}{(2\tau)^2}\sum_{n=1}^\infty c_n\, x^n $$
\begin{equation} \label{eq:q170} 
=-\frac{1}{(2\tau)^2}\sum_{n=0}^{\infty}c_{n+1}\,x^{n+1}
=\sum_{n=0}^{\infty}d_{n}\,x^{n+1},
\end{equation}
where the coefficients $d_n$ are 
\begin{equation}  \label{eq:q171} 
d_{n}=-\frac{1}{(2\tau)^2}\,c_{n+1}
=\frac{(-1)^n(2\tau)^n}{(n+1)!}\;f_{n}(\tau), 
\qquad\qquad n=0,1,2,\ldots
\end{equation}
with
\begin{equation} \label{eq:q172} 
f_{n}(\tau)=(-1)^{n+1}(2\tau)^n(n+1)!
\sum_{k=0}^{\infty}
\frac{2^{2n+2k+4}(2^{2n+2k+4}-1)B_{2n+2k+4}}{(2n+2k+2)!(2n+2k+4)}
\,{n+k+1\choose n+1}\,\tau^{2k}. 
\end{equation}
For $n=0$ one has 
\begin{equation} \label{eq:q173} 
f_{0}(\tau)=-\sum_{k=0}^{\infty}
\frac{2^{2k+4}(2^{2k+4}-1)B_{2k+4}}{(2k+2)!(2k+4)}
\,(k+1)\,\tau^{2k}
=\frac{\sinh\tau}{\tau\cosh^3\tau},
\end{equation}
where in the last step Eq. (\ref{eq:q159}) has been used. 
It is worth noticing that 
\begin{equation} \label{eq:q174} 
f_n(\tau)= (-1)^n \tau^n \left( {1\over \tau} {d \over d\tau}\right)^n 
f_0(\tau). 
\end{equation}
With the help of Eqs. (\ref{eq:q170}) and (\ref{eq:q171}), 
the spectrum of Eq. (\ref{eq:q161}) is put into the form 
\begin{equation} \label{eq:q175} 
{E\over A}
=\left(\frac{\tau^2 \, \cosh^2\tau}{\sinh^2\tau}\right)
\sum_{n=0}^{\infty}
\frac{(-1)^n(2\tau)^n}{(n+1)!}\;f_{n}(\tau)
\,(\ell(\ell+1))^{n+1},
\end{equation}
since $x=\ell(\ell+1)$ from Eq. (\ref{eq:q162}). 
It is clear that Eq. (\ref{eq:q175}) is an expansion in terms of 
$\ell(\ell+1)$, as expected. 

\section{Approximate Expansion of the su$_q$(2) ITO Spectrum} 

In the limit of $|\tau|<<1$ one is entitled to keep in Eq. (\ref{eq:q172}) 
only the term with $k=0$. Then the function $f_n(\tau)$ takes the form 
\begin{equation} \label{eq:q176} 
f_n(\tau) \rightarrow {(-1)^{n+1} 2^{2n+2} (2^{2n+4}-1)
B_{2n+4} \over (2n+1)!!  (n+2)}  \tau^n,
\end{equation}
where the Bernoulli numbers appear again and use of the identity 
\begin{equation} \label{eq:q177} 
(2n+2)! = 2^{n+1} (n+1)! (2n+1)!!
\end{equation} 
has been made. 
Taking into account the Taylor expansions 
\begin{equation} \label{eq:q178} 
\sinh x = x + {x^3\over 3!} + {x^5 \over 5!} + \cdots, \qquad
\cosh x = 1 + {x^2 \over 2!} + {x^4 \over 4!} + \cdots,
\end{equation}
and keeping only the lowest order terms, one easily sees that 
Eq. (\ref{eq:q175}) is put in the form 
\begin{equation} \label{eq:q179} 
{E\over A} \approx \sum_{n=0}^\infty  {2^{2n+4} 
(1-2^{2n+4}) B_{2n+4} \over (2n+2)! (2n+4)} (2\tau)^{2n}  (\ell(\ell+1))^{k+1},
\end{equation}
where use of the identity of Eq. (\ref{eq:q177}) has been made once more
and use of the fact that 
\begin{equation} \label{eq:q180}
{\tau^2 \cosh^2\tau \over \sinh^2\tau} \approx 1 
\quad {\rm for} \quad |\tau|<<1
\end{equation}
has been made.  
Comparing this result with Eq. (\ref{eq:q160})
and making the identifications
\begin{equation} \label{eq:q181} 
x=2\tau \sqrt{\ell (\ell+1)} = 2\tau \xi, \qquad \xi=\sqrt{\ell(\ell+1)},
\end{equation}
Eq. (\ref{eq:q179}) is put into the compact form
\begin{equation} \label{eq:q182} 
E\approx {A\over (2\tau)^2} \tanh^2(2\tau\sqrt{\ell(\ell+1)})
= {A\over (2\tau)^2} \tanh^2(2\tau\xi), \qquad q=e^\tau.
\end{equation}
The extended form of the Taylor expansion of $E$ is easily obtained from 
Eq. (\ref{eq:q179})
\begin{equation} \label{eq:q183} 
E\approx A \left( \ell(\ell+1)-{2\over 3} (2\tau)^2 (\ell(\ell+1))^2 
+{17\over 45} (2\tau)^4 (\ell(\ell+1))^3 -{62\over 315} (2\tau)^6 
(\ell(\ell+1))^4 +\cdots\right). 
\end{equation}
Eq. (\ref{eq:q182}) will be referred to as the ``hyperbolic tangent formula''. 

\section{Numerical Tests}

The formulae developed in the previous sections will be now tested against 
the experimental spectrum of the HF molecule, in which sizeable deviations 
from pure rotational behavior are observed. 
The data are taken
from the R branch ($\ell\to \ell+1$) and the P branch ($\ell\to \ell-1$) of 
Ref. \cite{HF}, using the equations \cite{Barrow}
\begin{equation}\label{eq:q184}
E^R_\ell-E^P_\ell= E_{\ell+1}(v=1)-E_{\ell-1}(v=1),
\end{equation}
\begin{equation}\label{eq:q185} 
E^R_\ell-E^P_{\ell+2}=E_{\ell+2}(v=0)-E_\ell(v=0),
\end{equation}
where $v$ is the vibrational quantum number.  
The purpose of this study is two-fold:

a) To test the quality of the approximations used in Secs. 9 and 13. 

b) To test the agreement between theoretical predictions 
and experimental data. 
 
The standard rotational expansion,
\begin{equation} \label{eq:q186}
E= A \ell(\ell+1) + B (\ell(\ell+1))^2 + C(\ell(\ell+1))^3 
+ D (\ell(\ell+1))^4 +\ldots,
\end{equation}
from which only the first two terms will be included in order to keep 
the number of parameters equal to two, 
as well as the Holmberg--Lipas two-parameter expression \cite{Lipas}
\begin{equation} \label{eq:q187}
E= a(\sqrt{1+b \ell(\ell+1)}-1),
\end{equation} 
which is known to give the best fits to experimental rotational nuclear
spectra among all two-parameter expressions \cite{Casten}, will be included 
in the test for comparison. 
For brevity we are going to use the following terminology: \hfill\break
Model I for Eq. (\ref{eq:q40}) (original su$_q$(2) formula), \hfill\break
Model I$'$ for Eq. (\ref{eq:q98}) (``the sinus formula''), \hfill\break
Model II for Eq. (\ref{eq:q143}) (``the su$_q$(2) irreducible tensor operator 
(ITO) formula''), \hfill\break
Model II$'$ for Eq. (\ref{eq:q182}) (``the hyperbolic tangent formula''), 
\hfill\break 
Model III for Eq. (\ref{eq:q186}) (the standand rotational formula), and 
\hfill\break 
Model IV for Eq. (\ref{eq:q187}) (the Holmberg--Lipas formula). \hfill\break
It should be emphasized at this point that in models I and I$'$ the
deformation parameter is a phase factor ($q=e^{i\tau}$, $\tau$ real), while 
in models II and II$'$ the deformation parameter is a real number 
($q=e^\tau$, $\tau$ real). A consequence of this fact is the presence 
of trigonometric functions in models I and I$'$, while in models II and II$'$
hyperbolic functions appear. 

The parameters resulting from the relevant least square fits, together with 
the quality measure 
\begin{equation} \label{eq:q188}
\sigma = \sqrt{ {2\over \ell_{max}} \sum_{i=2}^{\ell_{max}} 
(E_{exp}(\ell)-E_{th}(\ell))^2 },
\end{equation}
where $\ell_{max}$ is the angular momentum of the highest level 
included in the fit, are listed in Table 1, while in Table 2 the 
theoretical predictions of all models for the $v=0$ band of HF  
are listed together with the experimental spectrum. 

From these tables the following observations can be made: 

a) As seen in Tables 1 and 2, models I and I$'$ give results which are almost 
identical. The same is true for models II and II$'$. We therefore conclude 
that the approximations carried out in Secs. 5 and 10 are very accurate. 

b) Models II and II$'$ appear to give the best results, followed 
by models I and $I'$. Model III gives reasonable results, while 
model IV gives the worst ones, a result which comes as a slight 
surprise, since model IV is known to provide the best fits of 
nuclear rotational spectra \cite{Casten}. 

These observations lead to the following conclusions: 

a) One can freely use model I$'$ in the place of model I, and model II$'$
in the place of model II, since the relevant approximations turn out to be 
very accurate. Models I$'$ and II$'$ have the advantage of providing 
simple analytic expressions for the energy.

b) The fact that models II and II$'$ are better than models I and I$'$ 
indicates that within the same symmetry (su$_q$(2) in this case) it is 
possible to construct different rotational Hamiltonians characterized 
by different degrees of agreement with the data. 

c) The relative failure of Model IV for molecular rotational spectra
can be attributed to the fact that this model has been derived \cite{Lipas}
from a Bohr--Mottelson Hamiltonian \cite{BM}, which is certainly very 
appropriate for describing nuclear rotational spectra, but it is not 
necessarily equally appropriate for the description of molecular rotational 
spectra.  

\section{Discussion}

The main results of the present work are the following:  

a) The rotational invariance of the original su$_q$(2) Hamiltonian 
\cite{CPL175,RRS,PLB251}
under the usual physical angular momentum has been 
proved explicitly and its connection to the formalism of Amal'sky \cite{Amal}
(``the sinus formula'') has been given. 

b) An irreducible tensor operator (ITO) of rank one under su$_q$(2) has 
been found and used, through $q$-deformed tensor product and $q$-deformed 
Clebsch--Gordan coefficient techniques \cite{STK593,STK690,STK1068,STK1599},  
for the construction of a new Hamiltonian appropriate 
for the description of rotational spectra, the su$_q$(2) ITO Hamiltonian. 
The rotational invariance of this new Hamiltonian under the usual physical 
angular momentum has been proved explicitly. Furthermore, an approximate
simple closed expression (``the hyperbolic tangent formula'') 
for the energy spectrum of this Hamiltonian has been found.

From the results of the present work it is clear that the su$_q$(2) 
Hamiltonian, as well as the su$_q$(2) ITO Hamiltonian, are complicated 
functions of the Casimir operator of the usual su(2), i.e. of the square
of the usual physical angular momentum. These complicated functions 
possess the su$_q$(2) symmetry, in addition to the usual su(2) symmetry. 
Matrix elements of these functions can be readily calculated in the deformed 
basis, but also in the usual physical basis. A similar study of a
$q$-deformed quadrupole operator is called for. 

\section*{ACKNOWLEDGEMENTS}
The authors acknowledge support from the Bulgarian Ministry of Science 
and Education under Contracts No. $\Phi$-415 and $\Phi$-547.

\newpage 

\begin{table}

\caption{Parameter values and quality measure $\sigma$ (Eq. (\ref{eq:q188}))
for models I (Eq. (\ref{eq:q40})), I$'$ (Eq. (\ref{eq:q98})), II (Eq. 
(\ref{eq:q143})), II$'$ (Eq. (\ref{eq:q182})), III (Eq. (\ref{eq:q186})), 
and IV (Eq. (\ref{eq:q187})), obtained from least square fits 
(shown in Table 2) to the 
experimental spectrum of the $v=0$ band of HF, taken from Ref. \cite{HF}
through Eqs. (\ref{eq:q184}), (\ref{eq:q185}). 
}

\bigskip

\centering
\begin{tabular}{ r r r r r | r r | r r }
\hline
\hline 
I,I$'$,II,II$'$  &  I     & I$'$   & II     & II$'$ & III & III & IV & IV \\
$A$ (cm$^{-1}$) & 20.553 & 20.554 & 20.559 & 20.559 & $A$ (cm$^{-1}$) &
20.550 & $a$ (cm$^{-1}$) & 93982  \\
$10^2 \tau$    &  1.742 &  1.742 &  0.623 &  0.623 & $10^2 B$ (cm$^{-1}$) &
0.204 & $10^3 b$ & 0.438 \\
$\sigma$ (cm$^{-1}$)&  0.072 &  0.072 &  0.048 &  0.051 & $\sigma$ (cm$^{-1}$)
& 0.163 & $\sigma$ (cm$^{-1}$) & 0.313 \\
\hline
\end{tabular}
\end{table}

\newpage 

\begin{table}

\caption{Theoretical predictions of models I (Eq. (\ref{eq:q40})), 
I$'$ (Eq. (\ref{eq:q98})), II (Eq. (\ref{eq:q143})), 
II$'$ (Eq. (\ref{eq:q182})), III (Eq. (\ref{eq:q186})), 
and IV (Eq. (\ref{eq:q187})), obtained from least square fits to 
the experimental spectrum (exp.) of the $v=0$ band of HF, taken from Ref. 
\cite{HF} through Eqs. (\ref{eq:q184}), (\ref{eq:q185}).  
All energies are given in cm$^{-1}$. The relevant model parameters and 
quality measure $\sigma$ (Eq. (\ref{eq:q188})) are given in Table 1. 
}

\bigskip

\centering
\begin{tabular}{ r r r r r r r r  }
\hline
\hline 
$\ell$ & exp. &  I   & I$'$  & II    & II$'$ & III   &  IV   \\
 2 & 123.33 & 123.25 & 123.25 & 123.29 & 123.27 & 123.23 & 123.34 \\ 
 4 & 410.34 & 410.25 & 410.25 & 410.35 & 410.32 & 410.18 & 410.52 \\ 
 6 & 859.69 & 859.60 & 859.60 & 859.73 & 859.72 & 859.49 & 860.04 \\ 
 8 & 1469.2 & 1469.1 & 1469.1 & 1469.3 & 1469.3 & 1469.0 & 1469.6 \\ 
10 & 2235.9 & 2235.9 & 2235.9 & 2236.0 & 2236.0 & 2235.8 & 2236.2 \\ 
12 & 3156.1 & 3156.1 & 3156.1 & 3156.1 & 3156.1 & 3156.1 & 3156.1 \\
14 & 4225.3 & 4225.4 & 4225.4 & 4225.3 & 4225.2 & 4225.5 & 4224.9 \\
16 & 5438.4 & 5438.5 & 5438.5 & 5438.3 & 5438.3 & 5438.6 & 5438.0 \\
18 & 6789.6 & 6789.5 & 6789.5 & 6789.6 & 6789.7 & 6789.4 & 6790.0 \\
\hline
\end{tabular}
\end{table}

\end{document}